\documentclass[aps,pra,reprint,superscriptaddress,10pt,showpacs,a4paper,nofootinbib]{revtex4-1}
\usepackage{amsmath,amssymb,amsfonts}
\usepackage{mathtools}
\usepackage{graphicx}
\usepackage{txfonts}
\usepackage{color}

\usepackage[unicode,breaklinks]{hyperref}
\hypersetup{
    unicode=true,
    a4paper=true,
    plainpages=false,
    pdftitle={Spectral energy transport in two-dimensional quantum vortex dynamics},
    pdfauthor={T. P. Billam, M. T. Reeves, and A. S. Bradley},
    pdfsubject={Spectral energy transport in two-dimensional quantum vortex dynamics},
    colorlinks=true,
    linkcolor=blue,
    citecolor=blue,
    filecolor=black,
    urlcolor=blue
}
\newcommand{\eqnrefp}[1]{{[Eq.~(\ref{#1})]}}

\newcommand{\eqnreft}[1]{{Eq.~(\ref{#1})}}

\newcommand{\figreft}[2]{Fig.~\ref{#1}#2}

\newcommand{\figreftfull}[2]{Figure~\ref{#1}#2}
\newcommand{\figrefp}[2]{[Fig.~\ref{#1}#2]}

\newcommand{\rr}{\mathbf{r}}
\newcommand{\vv}{\mathbf{v}}
\newcommand{\kk}{\mathbf{k}}
\newcommand{\ww}{\mathbf{w}}
\newcommand{\uu}{\mathbf{u}}

\newcommand{\TT}{\tilde{T}}
\newcommand{\PP}{\tilde{\Pi}}

\begin{document}
\title{Spectral energy transport in two-dimensional quantum vortex dynamics}

\author{T. P. Billam}
\thanks{Author to whom correspondence should be addressed}
\email[Email: ]{thomas.billam@durham.ac.uk}
\affiliation{Jack Dodd Centre for Quantum Technology, Department of Physics, University of Otago, Dunedin 9016, New Zealand}
\affiliation{Joint Quantum Centre (JQC) Durham--Newcastle, Department of Physics, Durham University, Durham, DH1 3LE, UK}
\author{M. T. Reeves}
\affiliation{Jack Dodd Centre for Quantum Technology, Department of Physics,
University of Otago, Dunedin 9016, New Zealand}
\author{A. S. Bradley}
\thanks{Author to whom correspondence should be addressed}
\email[Email: ]{abradley@physics.otago.ac.nz}
\affiliation{Jack Dodd Centre for Quantum Technology, Department of Physics,
University of Otago, Dunedin 9016, New Zealand}

\date{\today}

\pacs{
03.75.Lm     
67.85.De     
47.27.-i     
}

\begin{abstract}
We explore the possible regimes of decaying two-dimensional quantum turbulence,
and elucidate the nature of spectral energy transport by introducing a
dissipative point-vortex model with phenomenological vortex-sound interactions.
The model is valid for a large system with weak dissipation, and also for
systems with strong dissipation, and allows us to extract a meaningful and
unambiguous spectral energy flux associated with quantum vortex motion.  For
weak dissipation and large system size we find a regime of hydrodynamic vortex
turbulence in which energy is transported to large spatial scales, resembling
the phenomenology of the transient inverse cascade observed in decaying
turbulence in classical incompressible fluids.  For strong dissipation the
vortex dynamics are dominated by dipole recombination and exhibit no
appreciable spectral transport of energy.
\end{abstract}

\maketitle
\section{Introduction}
Classical two-dimensional (2D) turbulence exhibits a well-established universal
phenomenology of spectral scaling laws and energy transport, where energy is
conservatively transported to large scales through a lossless inertial range, a
process known as the inverse energy cascade~\cite{Kraichnan1980a,
Les2008.Turbulence}.  While the Kolmogorov $-5/3$ power law associated with an
inertial range has been observed in simulations of two-dimensional quantum
turbulence (2DQT) in the context of dilute gas Bose-Einstein condensates
(BECs)~\cite{Neely2013a,Reeves2013a},  the nature of spectral energy transport
in 2DQT remains a crucial open question, with recent studies conflicting over
whether energy is transported to large \cite{Reeves2013a, Billam2014a,
Simula2014a} or small \cite{Chesler2013a, Numasato10b} scales.  As a model of a
compressible superfluid, the 2D Gross-Pitaevskii equation (GPE) can
simultaneously support several regimes of turbulence, including weak wave
turbulence~\cite{Proment09a,Nazarenko06a,Nazarenko07a}, negative temperature
point-vortex states in decaying \cite{Billam2014a, Simula2014a} and
forced~\cite{Bradley2012a,Reeves12a,Reeves2013a} quantum turbulence, and
evolution near non-thermal fixed points \cite{Nowak2011a, Nowak2012a,
Schole2012a}, offering a rich phenomenology involving coupled quantum vortex
and classical wave degrees of freedom.

The major challenge for understanding spectral energy transport in 2DQT is the
difficulty in identifying an unambiguous measure of spectral energy flux in a
superfluid that is compressible and, in general, dissipative
\cite{Numasato10a,Numasato10b}. Previous identifications of turbulent cascades
have relied on a combination of a spectral $k^{-5/3}$ scaling region and an
indirect measure of energy flux, such as vortex clustering and spectral energy
condensation~\cite{Reeves2013a}, energy flux through black-hole horizons in a
holographic gravity dual~\cite{Chesler2013a}, or other approximate methods
\cite{Numasato10b}, all of which are unable to unambiguously distinguish the
conservative transfer of vortex energy between spatial scales from the loss of
vortex energy due to the coupling to compressible degrees of freedom.
Furthermore, 2DQT studies have addressed systems for which the importance of
vortex-sound interactions varies greatly, leading to quite different results.
This diversity motivates a systematic exploration of the important physical
processes in regimes of 2DQT.

Here, we identify regimes of decaying 2DQT within GPE theory by varying the
system size and dissipation rate for a given initial condition, allowing
systematic control over the importance of vortex-sound interactions.  To study
the role of energy transport in the different regimes we introduce a
dissipative point-vortex model for quantum vortices, to which we add
phenomenological vortex-sound interactions. This model allows us to extract an
unambiguous spectral energy flux and analyze the effects of dissipation. The
model agrees well with the damped, projected Gross-Pitaevskii equation (dPGPE),
a physically well-justified model of BECs~\cite{Blakie08a}, both for large
system sizes and for strong dissipation. We find that for large system sizes
and weak dissipation relevant to BEC experiments \cite{Bradley2012a}, as
explored in Ref.~\cite{Billam2014a}, energy is indeed transported to large
scales by vortex interactions. We find time-averaged spectral characteristics
broadly consistent with the phenomenology of decaying turbulence in a classical
incompressible fluid \cite{Mininni2013a}, including an approximately lossless
inertial range with negative spectral energy flux. We term this
phenomenological regime \textit{hydrodynamic vortex turbulence}. 

In the cases of strong dissipation, and weak dissipation in a small system,
this phenomenology is replaced by a dissipative collapse of the spectrum,
driven by the formation of vortex dipoles and rapid vortex-antivortex
annihilation.  While our effective point-vortex model does not accurately
describe the dPGPE dynamics in the small system with weak dissipation (where
dominant vortex-sound interactions place the system in a strong wave-turbulent
regime), the long-time evolution in all these cases appears similar to dynamics
generated by the non-thermal fixed point identified in previous classical field
studies \cite{Nowak2011a, Nowak2012a, Schole2012a}.  This corresponds to a
low-energy, long-lived configuration of a small number of vortices, arranged in
vortex dipoles. In contrast to the regime of hydrodynamic vortex turbulence,
the vortex dynamics in these cases is found to lack any significant spectral
transport of energy through scale space, and has no direct analogue in
classical fluid turbulence.

\section{Gross--Pitaevskii model}
\begin{figure*}
\centering
\includegraphics[width=\textwidth]{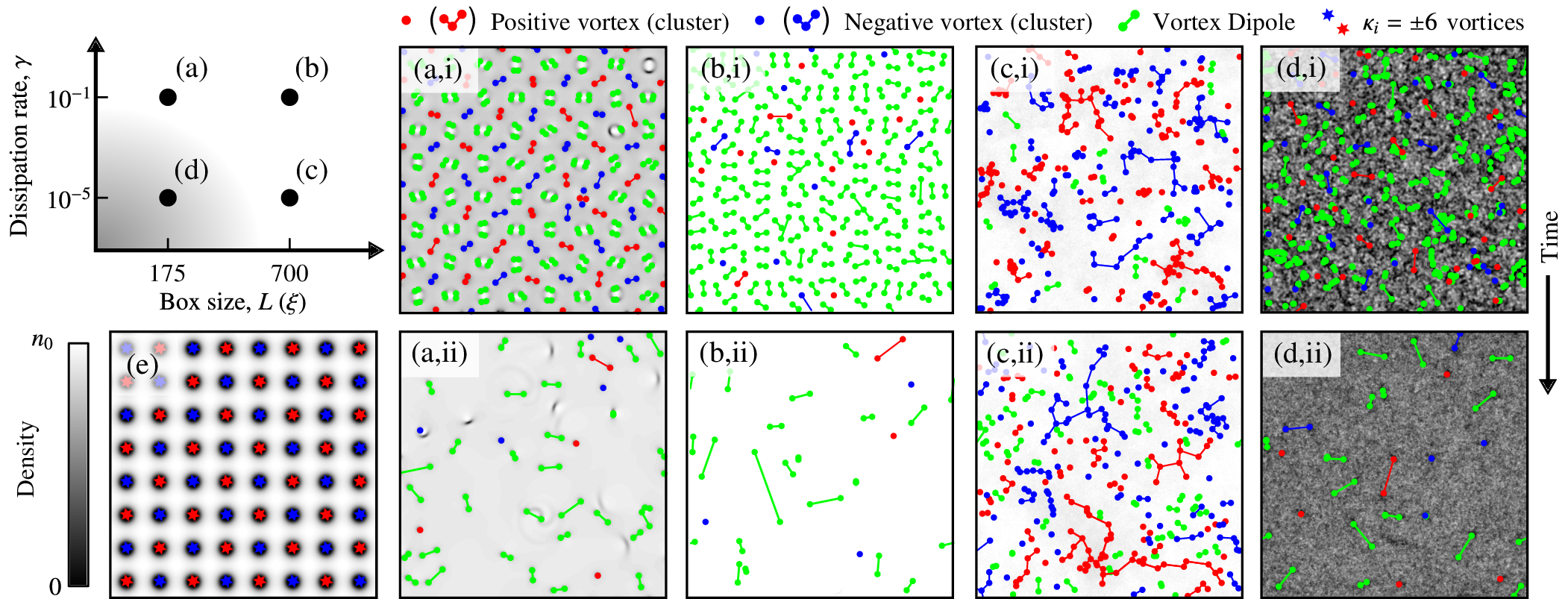}
\caption{(Color online) Regimes of decaying 2DQT following the decay of an unstable lattice of
charge $\kappa_i\pm6$ vortices, computed using the dPGPE (see also
\cite{Movies}). We show schematically the effects of dimensionless dissipation
rate, $\gamma$, and box size, $L$, (see axes, top left) on the dynamics: We
show the density (grayscale, background) and vortex configuration analyzed with
the recursive cluster algorithm \cite{Reeves2013a} (see key) at both short (i)
and long (ii) times after the break-up of the lattice, compared to the
characteristic timescale of the dynamics. In (e) we show the initial lattice
for $L=175$. The regimes can be characterised as dissipative dipole gas
dynamics (a,b), hydrodynamic vortex turbulence with energy transport to large
scales (c), and strong wave turbulence with close vortex-sound coupling (d)
[indicated schematically on axes (top left) by shaded
region].\label{fig:schematic}}
\end{figure*}

We model a finite-temperature BEC using the two-dimensional dPGPE~\cite{Blakie08a}  
\begin{equation}
i\hbar \frac{\partial \psi(\mathbf{r},t)}{\partial t} = (1-i\gamma) \mathcal{P}_c\left\{ \left(
\mathcal{L}[\mathbf{r},\psi(\mathbf{r},t)] - \mu \right)
\psi(\mathbf{r},t) \right\} \,,\label{eqn:dgpe_units} 
\end{equation}
where $\mathcal{L}[\mathbf{r},\psi(\mathbf{r},t)] = -\hbar^2\nabla_\perp^2/2m +
g_2 |\psi(\mathbf{r},t)|^2$, we have assumed oscillator length $l_z =
\sqrt{\hbar/m\omega_z}$, corresponding to tight harmonic confinement in the
$z$-direction, and $g_2 = \sqrt{8\pi}\hbar^2 a_s/ml_z$, where $m$ is the atomic
mass, $a_s$ the $s$-wave scattering length, and $\mu$ the chemical potential.
The projector $\mathcal{P}_c$ ensures complete numerical dealiasing within our
pseudospectral integration method \cite{Blakie08b, Billam2014a}. The
dimensionless damping rate $\gamma$ describes thermal dissipative collisions
between condensate atoms and non-condensate atoms; this is an important
physical effect in atomic superfluids, and has proved useful as both a
qualitative and semi-quantitative model for 2DQT experiments \cite{Neely2013a,
rooney_etal_pra_2013, Stagg2014a}.

We enumerate the regimes of decaying 2DQT by using \eqnreft{eqn:dgpe_units} to
simulate the decay of a periodic lattice of unstable charge-six ($\kappa_i =
\pm 6$) vortices, similar to Ref.~\cite{Chesler2013a}. We work in a periodic
square box of length $L\xi$ where $\xi = \hbar / \sqrt{\mu m}$ is the healing
length, and vary the parameters $L$ and $\gamma$; representative results are
shown in \figreft{fig:schematic}{} \cite{Movies}. In all cases the $\kappa_i =
\pm 6$ vortices rapidly disintegrate, forming a negative-temperature vortex
state with high point-vortex energy \cite{Billam2014a}. This state subsequently
evolves in different ways, depending primarily on the system size and
dissipation rate.

In the case of high dissipation [$\gamma=10^{-1}$,
\figreft{fig:schematic}{(a,b)}] the strong dissipation rapidly removes
compressible energy for all box sizes $L$.  The system also loses large amounts
of vortex energy, and large numbers of vortices to pair annihilation, and
evolves toward a long-lived, low-energy configuration of relatively few
vortices.  For weaker, experimentally relevant, dissipation [$\gamma=10^{-5}$]
and a small box size $L=175$, the initial disintegration of the $\kappa_i = \pm
6$ vortices releases compressible kinetic energy comparable to the
incompressible kinetic energy carried by the vortices, and the resulting strong
coupling between sound and vortices leads to a regime of strong wave turbulence
\figrefp{fig:schematic}{(d,i)}. However, at later times the sound and vortex
degrees of freedom cease to interact as strongly, and we observe a long-lived
configuration of few vortices arranged as dipoles, similar to those observed
for high dissipation in cases (a) and (b). As discussed in
Ref.~\cite{campbell_oneil_jsp_1991}, these positive-temperature configurations
of few vortices have no direct analogue in developed classical fluid
turbulence.

For weaker dissipation applicable to BEC experiments [$\gamma=10^{-5}$] and
a \textit{large} box size [$L=700$] the initial incompressible kinetic energy
dominates the compressible kinetic energy released by disintegration of the
$\kappa_i = \pm 6$ vortices \figrefp{fig:schematic}{(c)}. The subsequent
high-energy vortex dynamics, largely unaffected by dissipation or vortex-sound
interactions, are qualitatively different to the other cases. In this regime of
\textit{hydrodynamic vortex turbulence} one appears to recover the transport of
kinetic energy to large scales expected in classical 2D turbulence
\cite{Kraichnan1980a, Les2008.Turbulence, Mininni2013a} and reported in
Refs.~\cite{Bradley2012a,Billam2014a,Reeves2014a,Simula2014a}.

\section{Dissipative point-vortex model}
The $N$-vortex solutions of \eqnreft{eqn:dgpe_units} with approximately uniform
background density and well-separated vortex cores can be described by a
point-vortex model \cite{Fetter1966a}. Here, we develop such a model in order
to gain significant insights into the general dynamical regimes outlined above.
Our model includes the effects of dissipation due to $\gamma$ directly, and the
effects of coupling to sound phenomenologically.  Note that from here on, we
will focus on cases (a), (b), and (c), as in case (d) strong density
fluctuations preclude a quantitative description of the early-time dynamics in
terms of point-vortices. The later evolution of case (d) may be describable in
terms of a point-vortex model with a phenomenolgically altered value of
$\gamma$, but we will not pursue this possibility here. 

In the presence of dissipation, the general equation of motion for the $i$th
vortex, located at $\rr_i$ and with charge $\kappa_i = \pm1$ (circulation
$\kappa_i h/m$), is \cite{tornkvist_shroder_prl_1997}
\begin{equation}
\frac{d \rr_i}{dt} = \frac{h}{m} \left( \nabla \theta|_{\rr = \rr_i} - \gamma \kappa_i \hat{\mathbf{e}}_3 \times \nabla \theta|_{\rr=\rr_i} \right) \,,
\end{equation}
where $\theta$ is the phase of the wavefunction, and $\hat{\mathbf{e}}_3$ is a
unit vector in the $z$-direction . This leads to the
dissipative point-vortex model
\begin{equation}
\frac{d \rr_i}{dt} = \vv_i + \ww_i;\;\; \vv_i = {\sideset{}{'}\sum_{j=1}^N} \vv_{i}^{(j)};\;\; \ww_i = -\gamma \kappa_i \hat{\mathbf{e}}_3 \times \vv_{i}, \label{eqn:vortex_eom}
\end{equation}
where the primed summation indicates omission of the term $j=i$, and we have
explicitly separated the velocity of the $i$th vortex into a Hamiltonian part,
$\vv_i$, and a dissipative (Peach--Koehler \cite{Peach1950a}) part, $\ww_i$.
Hence, $\vv_{i}^{(j)}$ represents the velocity of the $i$th vortex due to the
$j$th in a Hamiltonian point-vortex model with appropriate boundary conditions.
We consider a periodic square domain of size $L\xi \times L\xi$, in which case
\cite{weiss_mcwilliams_pf_1991}
\begin{equation}
\vv_{i}^{(j)} = 
  \frac{\pi c \kappa_j } {L}
  \sum_{m=-\infty}^{\infty}
  \left( \begin{array}{c}
    \frac{-\sin(y_{ij}^\prime)}{\cosh(x_{ij}^\prime-2\pi m) - \cos(y_{ij}^\prime)}\\
    \frac{\sin(x_{ij}^\prime)}{\cosh(y_{ij}^\prime-2\pi m) - \cos(x_{ij}^\prime)}
  \end{array} \right),
\end{equation}
where $(x_{ij},y_{ij}) \equiv \rr_{ij} \equiv \rr_i - \rr_j$,
$(x_{ij}^\prime,y_{ij}^\prime) = 2\pi (x_{ij},y_{ij}) / L\xi$, and
$c=\sqrt{\mu/m}$ is the speed of sound.

The coupling between incompressible and compressible energy in an atomic BEC
introduces physics not captured by \eqnreft{eqn:vortex_eom} primarily when
vortices approach close to, or within, core-overlap distances.  We therefore
phenomenologically account for these effects within our model in two ways:
Firstly, similarly to Refs.~\cite{Simula2014a,campbell_oneil_jsp_1991}, we
remove vortex dipoles in which the constituent vortices approach each other
within one healing length $\xi$. This mimics the vortex-antivortex
annihilations that can be a key feature of quantum turbulence in compressible
superfluids \cite{Numasato10a, Simula2014a, Kwon2014a}. Secondly, we modify the
dissipation $\gamma$ for vortices with nearby same-sign neighbours, replacing
$\gamma$ in
\eqnreft{eqn:vortex_eom} with
\begin{equation}
\gamma_i = \mathrm{max} \left( \exp\left[\ln(\gamma) \frac{r_{is} - r_1}{r_2-r_1} \right],\; \gamma \right),
\end{equation}
where $r_{is}$ is the distance to the nearest same-sign neighbour of vortex
$i$, $r_2 = 10\xi$, and $r_1 = \xi$. This phenomenologically accounts for
radiation of sound by rapidly-accelerating vortices, an effect for which an
approximate analytic treatment exists for an isolated vortex pair
\cite{pismen,vinen_prb_2001}, but would appear to be impractical for general
$N$-vortex configurations. In practice, we find that our phenomenological model
captures the important physics of radiation while remaining reasonably
insensitive to variations in $r_1$ and $r_2$ around our chosen values.

\begin{figure*}
\centering
\includegraphics[width=\textwidth]{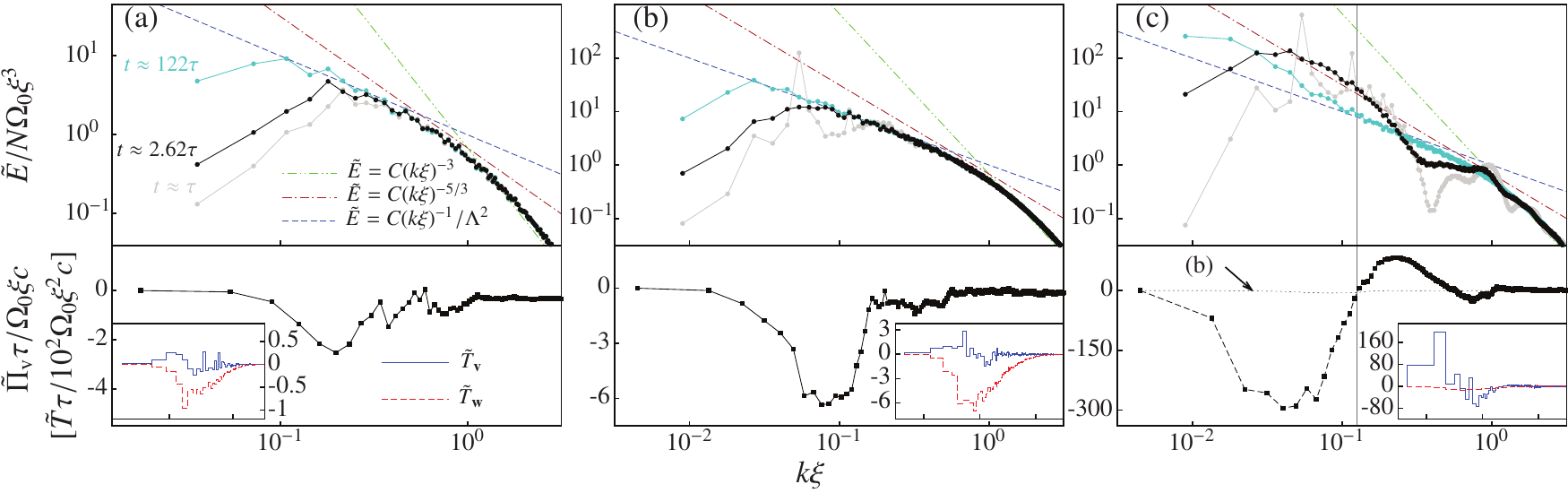}
\caption{(Color online) Spectral energy transport in the phenomenological
point-vortex model (see also supplementary movies \cite{Movies}): For the
parameters given in \figreft{fig:schematic}, (a--c) show ensemble and
time-averaged incompressible energy spectrum (top), flux (bottom), and transfer
function (bottom, inset).  Straight lines show the universal ultraviolet
spectrum $\tilde{E} = C(k\xi)^{-3}$ and the infrared spectrum for randomly
distributed vortices $\tilde{E} = C(k\xi)^{-1}/\Lambda^2$ (see text). A
Kolmogorov scaling law $\tilde{E} = C(k\xi)^{-5/3}$ is also shown for
reference.  Time derivatives are scaled to the turnover time $\tau$ (see text)
and time averaging is done over a window of duration $0.63\tau$ [$0.75\tau$] in
(a) [(b,c)], centered as indicated.  Transfer functions and fluxes are shown
for time $t\approx 2.62 \tau$. Vertical lines in (c) indicate the transition
from negative to positive flux.  The spectrum and flux in case (c) are broadly
consistent with results expected for classical decaying turbulence
\cite{Mininni2013a}.
\label{fig:flux}}
\end{figure*}

\section{Spectral analysis}
Using our phenomenological point-vortex model, we simulate ensembles of 10
trajectories in a scenario similar to \figreft{fig:schematic}{} for 15--20
turnover times $\tau \equiv L\xi/u_{\rm rms}$, where $u_{\rm rms}$ is the
average root-mean square velocity over the ensemble at $t=0$. The ensemble is
defined by beginning from configurations in which six appropriately- and
likewise-circulating vortices are randomly placed in a circle of radius $10\xi$
centered at the location of each lattice site, with an enforced minimum
separation of $\xi$ between vortices. This closely reproduces configurations
observed during disintegration of $\kappa_i = \pm 6$ vortices in the dPGPE
after short times. Because of the interaction between compressible degrees of
freedom and dissipation in the decay of a $\kappa_i = \pm 6$ vortex in the
dPGPE, the timescales over which the initially separated groups of vortices mix
are not completely consistent between the dPGPE and the point-vortex model
(that does not include these effects). For $\gamma=10^{-1}$ the dPGPE exhibits
faster mixing, while for $\gamma=10^{-5}$ the point-vortex model mixes faster.
However, once the initial clusters have mixed, and the vortices become
well-separated, the point-vortex and dPGPE vortex dynamics become similar in
all cases (a--c) \cite{Movies}.

The incompressible kinetic energy spectrum of our point-vortex model can be
written analytically in terms of the vortex positions, an approach pioneered by
Novikov \cite{Novikov1975} for the pure point-vortex model. Describing quantum
vortex cores by the ansatz $\psi(|\rr-\rr_i|,\phi) = n_{0}
[r^2/(r^2+\xi^2\Lambda^{-2})]^{1/2} e^{i\kappa_i \phi}$, where $n_{0}$ is the
background homogeneous superfluid density and $\Lambda \approx 0.825$
\cite{Bradley2012a}, in the periodic square box of length $L\xi$ the
incompressible kinetic energy spectrum is \cite{Billam2014a}
\begin{equation}
\frac{E^i(\mathbf{k})}{N} = G_\Lambda(k\xi) \left[ 1 + \frac{2}{N} \sum_{i=1}^{N-1} \sum_{j=i+1}^N \kappa_i \kappa_j \cos(\mathbf{k}\cdot\mathbf{r}_{ij}) \right], \label{eqn:spectrum}
\end{equation}
where $G_\Lambda(k\xi) = \Omega_0 \xi^4 g(k\xi/ \Lambda)/(\pi \Lambda k\xi)$,
$\Omega_0 = \pi \hbar^2 n_0/m\xi^2$ is the quantum of enstrophy, $g(z) = (z/4)
[ I_1(z/2) K_0(z/2) - I_0(z/2) K_1(z/2)]^2$ (for modified Bessel functions
$I_\alpha$ and $K_\alpha$), and the momentum $\kk = (n_x \Delta k, n_y \Delta
k)$, for $\Delta k = 2\pi /L \xi$ and $n_x,n_y \in \mathbb{Z}$.  As established
previously \cite{Bradley2012a}, \eqnreft{eqn:spectrum} leads to an
angularly-integrated kinetic energy spectrum, $E^i(k) = k\int_0^{2\pi}
E^i(\mathbf{k}) d\phi_k$, that exactly obeys the universal power-law $E^i(k) =
C(k\xi)^{-3}$ in the ultraviolet ($k\gg\xi^{-1}$), where $C= N\Lambda^2
{\Omega_0 \xi^3}$ and $k=|\mathbf{k}|$.  In the infrared ($k \lesssim
\xi^{-1}$) the spectrum of randomly distributed vortices obeys the power-law
$E^i(k) = C(k\xi)^{-1} / \Lambda^2$.

In order to study spectral energy transport in our model, we consider the
\textit{transfer function} of classical turbulence, $T(\kk) = dE^i(\kk)/dt$
\cite{Les2008.Turbulence}, given by
\begin{equation}
T(\kk) = -2 G_\Lambda(k\xi) \sum_{i=1}^{N-1} \sum_{j=i+1}^N \kappa_i \kappa_j \sin(\mathbf{k}\cdot\mathbf{r}_{ij}) \kk\cdot\left( \frac{d \rr_{i}}{dt} - \frac{d\rr_j}{dt}\right), \label{eqn:tk}
\end{equation}
where the vortex velocities are known from
\eqnreft{eqn:vortex_eom}.  One can split $T(\kk)$ into Hamiltonian and
dissipative parts $T(\kk) = T_\vv(\kk) + T_\ww(\kk)$, where
\begin{equation}
 T_\uu(\kk) = -2G_\Lambda(k\xi) \sum_{i=1}^{N-1} \sum_{j=i+1}^N \kappa_i \kappa_j \sin(\mathbf{k}\cdot\mathbf{r}_{ij}) \kk\cdot\left( \uu_i - \uu_j\right), \label{eqn:tku}
\end{equation}
for $\uu = \{\vv,\ww\}$. Since we are interested in spectral energy transport
at large scales, which may be strongly influenced by the box size in negative
temperature states \cite{campbell_oneil_jsp_1991}, we avoid approximating the
angularly-integrated transfer function $T(k) = \int_0^{2\pi} T(\kk) kd\phi_k$
analytically. Instead we introduce the discrete transfer function for $n =
1,2,\ldots$
\begin{equation}
\TT (n\Delta k) =  \sum_{\mathclap{(n-\frac{1}{2})\Delta k < |\kk| \le (n+\frac{1}{2}) \Delta k}} T(\kk) \,\Delta k, \label{eqn:acc}
\end{equation}
that measures the rate of change of energy in the wavenumber range specified
in the sum.
From \eqnreft{eqn:acc} we may define the discrete flux
for $n=0,1,\ldots$
\begin{equation}
\PP \left[ (n+1/2)\Delta k \right] = - \sum_{m=1}^n \TT(m,\Delta k)\, \Delta k.\label{eqn:boxflux}
\end{equation}
For $\gamma=0$, this defines the instantaneous energy flux through the
$k$-space surface $|\kk| = (n+\frac{1}{2})\Delta k$ \footnote{This becomes
equal to the conventional continuum flux $\Pi(k) = - \int_0^k T(k') dk'$ in the
high-$k$ limit, by setting $k = (n+\frac{1}{2})\Delta k$.}. In the presence of
dissipation the meaning of $\PP[(n+1/2)\Delta k]$ is not clear. However,
crucially, the partition of $T(\kk)$ into Hamiltonian and dissipative parts
\eqnrefp{eqn:tku} carries over naturally to $\TT(n \Delta k)$ and $\PP [(n+1/2)
\Delta k]$.  Hence, we are able to define three important quantities;
$\TT_\vv(n \Delta k)$, the change in energy due to energy transport to or from
other scales; $\TT_\ww(n\Delta k)$, the change in energy due to dissipative losses;
and $\PP_\vv[(n+1/2)\Delta k]$, the energy flux ignoring dissipative losses.

Numerically, evaluation of $\TT_\vv$, $\TT_\ww$, and $\PP_\vv$ up to and beyond
the vortex core momentum scale ($k\sim \xi^{-1}$) is expensive for large $N$
and large $L$.  To overcome this numerical challenge we developed GPU codes
\cite{cudaRef} that allow us to evaluate these quantities for many vortex
configurations ($\sim 10^4$) within a reasonable timeframe. When suitably
averaged over time and ensemble, in analogy to studies of classical turbulence
\cite{Mininni2013a}, these quantities offer a detailed picture of spectral
energy transport in our point-vortex model.

\section{Results and Discussion}
\figreftfull{fig:flux}{} and the supplementary movies \cite{Movies} provide
support our previous classification of the regimes of 2DQT, presented in
\figreft{fig:schematic}{}, and provide quantitative confirmation of the nature
of spectral energy transport in cases (a--c).

In the strongly dissipative cases \figrefp{fig:flux}{(a,b)} the relative
magnitude of $\TT_\vv$ and $\TT_\ww$ are comparable, highlighting the key role
of dissipation in this regime. The absence of any lossless inertial range rules
out the existence of a turbulent cascade. Instead the spectrum undergoes a
dissipative collapse: the total kinetic energy and the vortex number decrease
rapidly, with little energy transport occurring. Note that we plot the kinetic
energy spectrum $\tilde{E}(k)$ divided by the average number of vortices $N$ in
\figrefp{fig:flux}{} and the supplementary movies, in order to clearly show the
spectrum as a function of $k$ at all times. This can create the impression of a
transfer of energy towards low $k$ during the time-development of our plots of
the spectrum; however, this is caused by the fact that both dissipation and
vortex annihilation remove energy preferentially at high $k$. As indicated by
the plots of $\TT_\vv$, $\TT_\ww$, and $\PP_\vv$, the actual transfer of energy
to large scales due to vortex interactions is not significant.

For the case of hydrodynamic vortex turbulence \figrefp{fig:flux}{(c)},
dissipation is not significant. The spectrum and flux resemble those of
classical decaying 2D turbulence generated from a similar initial concentration
of energy, where a transient dual energy-enstrophy cascade is observed
\cite{Mininni2013a}. While the scale ranges are small (as expected for a system
of a few hundred vortices when compared to classical turbulence), the spectrum
does show some respective $k^{-5/3}$ and $k^{-3}$ character above and below the
scale of the transition from negative to positive flux (vertical line),
suggesting that a quasi-classical dual energy-enstrophy cascade may be an
emergent phenomenon in large-scale 2DQT. 

A potentially powerful feature of our point-vortex model, due to its
phenomenological treatment of effects below the vortex core scale, is that it
may also be useful to describe the large-scale behaviour of vortices in other
superfluids with different microscopic vortex core structure. For instance, as
we describe in detail in Appendix A, we suspect the regime of dissipative
collapse \figrefp{fig:flux}{(a,b)} may describe aspects of the holographic
superfluid investigated in Ref.~\cite{Chesler2013a}, in which a direct cascade
of energy was previously inferred. This offers a possible resolution of the
apparent conflict between Ref.~\cite{Chesler2013a} and Refs.~\cite{Reeves2013a,
Billam2014a, Simula2014a} over the direction of energy transport in decaying
2DQT.

\section{Conclusions}
Firstly, we used the damped, projected Gross--Pitaevskii equation (dPGPE) to
examine several regimes of decaying 2DQT resulting from an unstable
lattice-type initial condition. By varying the system size and the dissipation
rate we showed the existence of at least three regimes: dissipative dipole gas,
strong wave turbulence, and hydrodynamic vortex turbulence. These findings help
to clarify the relation between previous studies of decaying 2DQT that have
examined systems from different regimes. 

Secondly, we introduced a phenomenological point-vortex model that accurately
describes the vortex dynamics of decaying 2DQT in most cases (excepting the
case of strong wave turbulence). We have used this model to elucidate the
nature of spectral energy transport due to quantum vortex dynamics. We find
that significant spectral energy transport due to vortex interactions occurs
only for large systems subject to weak dissipation, and that in this case kinetic
energy is transported to large scales. Our results for both energy spectra and
energy fluxes provide some indication that a transient dual energy-enstrophy
cascade, analogous to that present in the decay of turbulence in a classical
fluid, may emerge in large-scale 2DQT. Future work will focus on understanding
this emergent phenomenon in both forced and decaying turbulence.

\acknowledgments
We thank B.~P.~Anderson, S.~A.~Gardiner, and R.~Gregory for valuable
discussions. We are grateful to P.~M.~Chesler and A.~Lucas for discussions
regarding Ref.~\cite{Chesler2014a}. This work was supported by The New Zealand
Marsden Fund, and a Rutherford Discovery Fellowship of the Royal Society of New
Zealand (ASB). TPB was partly supported by the UK EPSRC (EP/K030558/1). We
acknowledge the NZ eScience Infrastructure (http://www.nesi.org.nz) and the
University of Otago for providing access to Nvidia Tesla GPUs.

\appendix
\section{CONNECTIONS TO HOLOGRAPHIC SUPERFLUIDS}

\begin{figure}[h]
\centering
\includegraphics[width=\columnwidth]{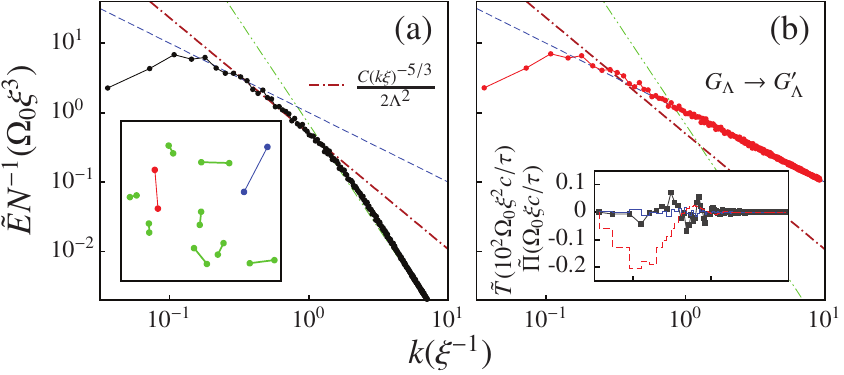}
\caption{(Color online) Accidental Kolmogorov $k^{-5/3}$-scaling close to the vortex-core
crossover: In (a), ensemble averaging over quiescent, non-turbulent, low-energy
configurations of 20 vortices consisting mostly of vortex dipoles ($L=175$, example inset) yields an appreciable
apparent $k^{-5/3}$-scaling region directly connected to the $k^{-3}$
vortex-core scaling. In (b), analytically removing the vortex-core scaling
destroys this $k^{-5/3}$ region , which for $\gamma=10^{-1}$ is clearly not related
to a lossless inertial range (see inset). Curves as in Fig. 2 of the main text.
\label{fig:dipoles}}
\end{figure}

In the main text, we have considered decaying two-dimensional quantum
turbulence in a compressible superfluid described by a Gross-Pitaevskii model.
Our results show that in vortex-dominated regimes where sound energy is
negligible [our cases (a), (b), and (c)] one observes either dissipative
collapse of the spectrum [strong dissipation, cases (a), (b)] or transport of
energy to large scales [weak dissipation, case (c)]. One question prompted by
these results is how they may be connected to turbulence in holographic 2D
superfluids \cite{Herzog2009a, Keranen2010a, Chesler2013a} for which a direct
cascade of energy was recently inferred in \cite{Chesler2013a}.

A problem in connecting our results and those of Ref.~\cite{Chesler2013a} is
that, unlike the Gross-Pitaevskii description, holographic superfluids are not
explicitly constructed to describe experimental atomic superfluids. Indeed, the
structure of vortex cores in holographic superfluids has been shown to be
different from the Gross-Pitaevskii form \cite{Keranen2010a}. Nonetheless, we
note that movies \cite{Movies} of the dPGPE evolution in case (a) of our work
are very similar to the movies of holographic superfluid vortex dynamics in
Ref.~\cite{Chesler2013a}.  In particular, both exhibit strong damping of sound
waves, rapid vortex-antivortex annihilation, and evolution towards a
positive-temperature vortex configuration.  This similarity suggests that the
holographic superfluid in Ref.~\cite{Chesler2013a} probes a regime of small
system size (relative to the vortex core) with dissipation several orders of
magnitude stronger than realized in atomic BEC experiments --- where typically
$\gamma \sim 10^{-5}$ -- $10^{-4}$ \cite{Bradley2012a}.  The possibility of
describing the holographic superfluid of Ref.~\cite{Chesler2013a} using a
point-vortex model with large dissipation was also recently pointed out
elsewhere \cite{Chesler2014a}.  If such a description is appropriate then,
according to our results, one would anticipate a dissipative collapse of the
spectrum in the scenario of Ref.~\cite{Chesler2013a}, rather than a $k^{-5/3}$
spectrum and a direct cascade of energy.

Despite the possibility of significant differences between
Gross-Pitaevskii-like and holographic superfluids, for a state with negligible
density fluctuations outside of the vortex cores, the kinetic energy spectra of
Gross-Pitaevskii-like and holographic superfluids should in fact be very
similar on physical grounds. In particular, in the ultraviolet the linear
scaling of the wavefunction at the vortex core in both superfluids
\cite{Keranen2010a} yields a $k^{-3}$ spectrum, while in the infra-red both
systems must recover a configurational point-vortex spectrum. Thus, differences
between the Gross-Pitaevskii and holographic kinetic energy spectra should be
confined to the crossover region ($k \sim \xi^{-1}$). Consequently, a
$k^{-5/3}$ spectrum in the configurational spectrum of a holographic superfluid
should be understandable in terms of a point-vortex model.

In \figreft{fig:dipoles}{(a)} we show the average incompressible kinetic energy
spectrum of 1000 random neutral 20-vortex configurations with point-vortex
energy (per vortex) $\approx -1$ (see Ref.~\cite{Billam2014a}). Such
configurations [example inset in \figreft{fig:dipoles}{(a)}], which do not
correspond to classical fluid turbulence \cite{campbell_oneil_jsp_1991},
resemble configurations obtained in case (a) and Ref.~\cite{Chesler2013a} after
most vortices have been annihilated in a dissipative collapse. Nonetheless, the
average spectrum exhibits a roughly half-decade sized region with apparent
$k^{-5/3}$ scaling \figrefp{fig:dipoles}{(a)}. However, replacing
$G_\Lambda(k\xi) \rightarrow G'_\Lambda(k\xi) = \Omega_0 \xi^4 / \pi(k\xi)^2$
allows recovery of the pure point-vortex spectrum \figrefp{fig:dipoles}{(b)}
for the same vortex configuration \cite{Bradley2012a}.  This spectrum follows
the single-vortex $k^{-1}$ scaling throughout the apparent $k^{-5/3}$ region in
\figreft{fig:dipoles}{(a)}, showing that the latter is due to the vortex core
structure (contained in $G_\Lambda$) and is unrelated to vortex interactions
and turbulence. Consistent with this picture, the ensemble-averaged transfer
function and incompressible energy flux [inset in \figreft{fig:dipoles}{(b)}]
show no lossless inertial range. 

While the vortex core structure in a holographic superfluid may be different in
its details, it quite likely leads to a similar ``accidental'' $k^{-5/3}$
scaling.  This offers an alternative explanation for the $k^{-5/3}$ scaling in
the spectrum attributed to a direct cascade in Ref.~\cite{Chesler2013a}. An
intriguing question that arises is whether adjusting parameters in the
holographic description might yield a regime of weakly-dissipative holographic
superfluid behaviour similar to our case (c), recovering the statistical
transport of energy to large scales associated with classical 2D turbulence.
Recent work \cite{Chesler2014a} has suggested a positive answer to this
question by analogy to a weakly-dissipative point-vortex model, but this has
not presently been confirmed by direct simulation of a weakly-dissipative
holographic superfluid.  Directly addressing this question may help to further
elucidate the physical nature of holographic superfluids.

%

\end{document}